# The order-theoretical foundation for data flow security

Luigi Logrippo [1,2]

[1]Département d'informatique et ingénierie, Université du Québec en Outaouais, Gatineau, Québec, CANADA J8X 3X7
[2]School of Electrical Engineering and Computer Science, University of Ottawa, Ottawa, Ontario, CANADA K1N 6N5

logrippol@acm.org

**Abstract.** Some theories on data flow security are based on order-theoretical concepts, most commonly on lattice concepts. This paper presents a correspondence between security concepts and partial order concepts, by which the former become an application of the latter. The formalization involves concepts of data flow, equivalence classes of entities that can access the same data, and labels. Efficient, well-known algorithms to obtain one of these from one of the others are presented. Security concepts such as secrecy (also called confidentiality), integrity and conflict can be expressed in this theory. Further, it is shown that complex tuple labels used in the literature to express security levels can be translated into equivalent set labels. A consequence is that any network's data flow or access control relationships can be defined by assigning simple set labels to the entities. Finally, it is shown how several partial orders can be combined when different data flows must coexist.

## 1. Introduction

Many books and papers on data security quickly introduce some order-theoretical concepts without fully justifying their use, and this is also the case for our paper [14]. The purpose of this paper is to show that there is a precise correspondence between basic concepts in data security theory and concepts found in the theory of partial orders. The latter also offers efficient algorithms. With these insights, some established concepts and their consequences become simpler and clearer, hopefully opening the door to further results. In particular, we obtain a generalization of the lattice-based theory pioneered by Bell and La Padula [3], Denning [9] and Sandhu [16] and developed by many others.

While the literature on the use of the lattice concept in data security theory is vast, there is almost no literature on the theoretical subject of this paper, in any case we refer to [14] for additional references.

## 2. The data flow relation and the partial order of equivalence classes

**Definition 1**: A *network N* is a finite set of entities with a binary relation *Channel*. Each entity has a unique *Name*.

Letters *x, y, z* will be used for variables over entities.

**Definition 2.** The binary relation *CanFlow* (written *CF*) is the reflexive, transitive closure of the relation *Channel*.

We use the following standard definitions: A *preorder* (also called quasi-order) is a reflexive, transitive relation. A *partial order* is a reflexive, transitive, antisymmetric relation. Thus *CF* is a preorder relation. *Channel* is an arbitrary relation.

**Definition 3.** Entities *x* and *y* are *equivalent* if *CF(x,y)* and *CF(y,x)*. An equivalence class of entities including *x,y,....* is denoted *[x,y,...]*.

**Lemma 1.** The relation between equivalence classes of entities is a partial order, i.e. a reflexive, transitive, antisymmetric relation, which we denote ⊑. Furthermore, *CF(x,y)* iff *[x]⊑[y]*.
*Proof:* Proofs of this result can be found in the literature, both order-theoretic and graph-theoretic. See for proofs and examples Chapt. 1, Th. 3 in Birkhoff [5], §2 in Fraïssé [10] and, for similar results in graph theory, Chapt. 3 in Harary et al. [11], Sect. 1.5 in Bang Jensen, Gutin [2]. The intuitive reason is that, by collapsing equivalent entities in a preorder into a single entity, the relation becomes antisymmetric and thus is a partial order.

Following common terminology, partially ordered sets will be called *posets*. Thus the set of equivalence classes of entities with the relation ⊑ is a poset. *[x]⊑[y]* is often expressed by saying that *[y] dominates [x]*. We also use the concept of *strict domination,* where *[x]⊑[y]* but *[x]≠[y]*. Posets have top (bottom) elements, i.e. elements that are not strictly dominated by any others (do not strictly dominate any others).

### 3. Labels

**Definition 4.** Let *Names* be the set of all names of entities in a network. We associate with each equivalence class *[x]* a set, called *Lab([x]),* which is a subset of *Names.* For each *[x]*, let *Ownlabel([x]) = {Name(y) | y∈[x]}*. For each *[x]*, let *Lab([x]) = ∪{Ownlabel([y]) | [y] ⊑ [x]}*.

**Algorithm 1 (to calculate labels from the *CF* relation).** The previous definition suggests the following linear-time algorithm to calculate the labels of entities for our finite networks.

- **Step 1.** Starting from the *CF* relation, the partial order of equivalence classes in a network is calculated by using a classical strongly connected component algorithm, as described in Stambouli, Logrippo [18]. Such algorithms have linear-time complexity (Sect. 20.5 in Cormen et al. [8]).
- **Step 2.** If *[x]* is a bottom equivalence class in the partial order, *Lab([x]) = Ownlabel([x])*. For all other *[x]*, the labels can be computed after the labels of all the equivalence classes that are strictly dominated by *[x]* have been computed (see Fig.1). This is also a linear-time construction.

The following result is true for all posets. In its generic formulation it says that *any partially ordered set is isomorphic to a subset of a power set, ordered by the subset relation.* This result is considered to be so elementary and basic in order theory that is seldom mentioned and not formally proved (Harzheim [12], Preface). We formulate and prove it for our finite networks as follows:

**Lemma 2:** The partial order relation between sets of equivalent entities is isomorphic to the set of labels ordered by the subset relation.

*Proof.* To see that the relation is one-to-one, by Def. 4 for each equivalence class there is a label. Since the set of entities in the different equivalence classes are different, then it must be that the the *Ownlabels* of these classes are also different. Hence the labels of two different equivalence classes, in which the *Ownlabels* are included, must also be different. On the other hand, it is impossible by construction that two different labels are assigned to any equivalence class. By the definition of label, it is also clear that we have *[x]⊑[y]* iff *Lab([x])⊆Lab([y])*.

**Definition 5.** For an entity *x*, we take *Lab(x)=Lab([x])*.

### 4. Basic results

**Theorem 1.** *CF(x,y) iff [x]⊑[y] iff Lab(x)⊆Lab(y).*
*Proof.* From Lemma 1 and Lemma 2 with Def. 5.

**Theorem 2.** For a set of entities, given any of: a *Channel* relation in the set, a *CF* relation, a partial order of equivalence classes of entities in the set, or a set of labels of entities in the set, the other three, satisfying Theorem 1, can be calculated with linear time or polynomial-time algorithms.

*Proof:*
i) Given a *Channel* relation, the *CF* relation can be computed by using transitive closure algorithms, see Sect. 23.2 in Cormen et al [8]. These algorithms have cubic, which is polynomial, complexity.
ii) Given a *CF* relation, its partial order of equivalence classes can be computed by using the mentioned strongly connected component algorithms (see Step 1 of Algorithm1), with linear time complexity.
iii) Given a partial order of equivalence classes, the labels of the equivalence classes and entities can be computed by using the linear-time construction of of Algorithm 1.
iv) Given a set of labeled entities, the *CF* relation or the partial order of equivalence classes can be computed by checking for inclusion among pairs of labels. Set inclusion is a problem having the same complexity as sorting, which is linearithmic (Ben-Or [4]).
v) A *Channel* relations can be calculated from a *CF* relation, most trivially by defining *Channel(x,y) = CF(x,y)*. This will give all possible channels. Reduced *Channel* relations can then be computed by transitive reduction algorithms, of cubic complexity (Aho et al. [1]); however this might remove channels that could be useful for implementation, see Sect. 5.

## 5. Example with data security concepts

Intuitively, *Channel(x,y)* should be taken to mean that data can move from entity *x* to entity *y*. So *Channel* denotes an authorization, permission or right and not the execution of an operation. It can denote an access control permission (a *true* value in an access control matrix [13]) or a possibility of reading or writing data by the use of encryption-decryption methods. Many methods exist to specify *Channel* relations. When all variables are fixed, a DAC system, a RBAC system, an ABAC system, a routing table or a set of entities communicating by encryption and decryption all implicitly define access control matrices [18]. So, in term of actual operations, *Channel(x,y)* can denote an authorization of *x writing* on *y*, *y reading* from *x*, *x sending* to *y*, *y receiving* from *x*, or similar for *pushing* and *pulling*, *putting* and *getting*, etc. An equivalence class is a set of entities that are authorized to share all data. Labels identify the data categories to which entities have access, or their data's *provenance.*

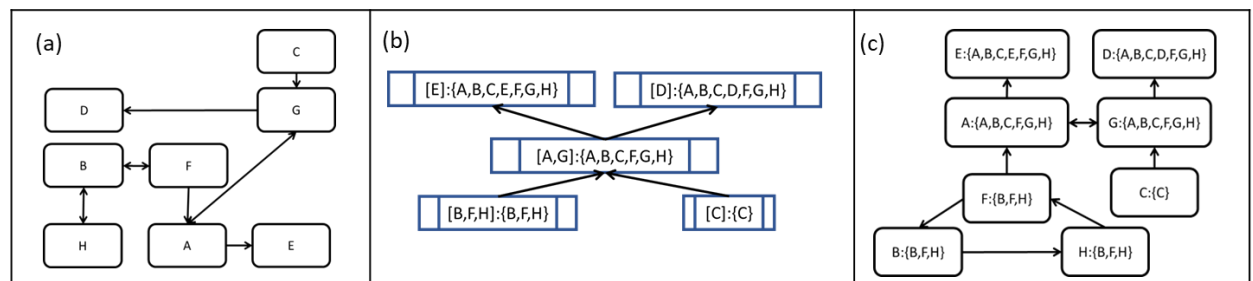

**Fig. 1.** A *Channel* relation (a), its labeled equivalence classes (b), and labeled entities (c).

In the diagrams of Fig. 1 we see:
- in (a) an arbitrary *Channel* relation among entities;

- in (b) the corresponding partial order ⊑ of labeled equivalence classes of entities (Lemma 1), with labels (Def. 4) preceded by colons;
- in (c) an assignment of labels to entities (Def. 5), with labels preceded by colons, and with a reduced *Channel* relation consistent with (b).

The order relations are shown in (b) and (c) with greater elements above smaller ones. For readability, they are shown transitively reduced and without reflexive edges.

Note the following:

- The following non-trivial equivalence classes were detected: *[B,F,H]* and *[A,G]*.
- By Th.1, all diagrams of Fig. 1 specify the same data flow. In particular, the labels in (b) and (c) define the data flow in (a).
- In established theory, *CF(x,y) iff Lab(x)⊆Lab(y)* characterizes Multi-level systems within a lattice framework; it is true for any network in our theory.
- If we define a *maximum secrecy* (also called *top secrecy* or *top confidentiality*) equivalence class of entities to be an equivalence class that has no outgoing data flows (that it is not strictly dominated by any other), we see two maximum secrecy equivalence classes, *[D]* and *[E]*. The label of these entities in not included in any other.
- If we define a *maximum integrity* equivalence class of entities to be an equivalence class of entities that has no incoming data flows (that does not strictly dominate any other) (Sect. IV in Bell, La Padula [3], or Sect. 4 in Sandhu [17]), we see two maximum integrity equivalence classes of entities, *[B,F,H]* and *[C]*. The label of these equivalence classes is simply their *Ownlabel*.
- If we define two entities to be in *conflict* if there is no entity to which all of them can flow, then entities *E* and *D* are in conflict.
- A partial order of equivalence classes of entities such as (b) defines an equivalence class of *Channel* relations. Both (a) and (c) belong to this equivalence class, or can be said to be *implementations* of (b). Some channels that are not shown in (a) or (c) could be useful in practice, however, depending on implementation constraints on the physical network on which the *CF* relation should be implemented and on other constraints such as channel speed. The reader may be interested in discovering all implementations of (b) as an exercise. As mentioned, by taking *Channel(x,y)=CF(x,y)* we would have an implementation with all possible channels, which could be used if channels have no cost.

## 6. Translation of tuple labels to set labels

So far, we have considered only labels that are simple sets of categories. In security, there is often consideration of tuple labels, representing different security properties, such as secrecy or integrity levels, as well as data categories. For example, in Bishop[6], Ch. 5, a label *<SECRET, {EUR,US}>* can apply to entities containing data of secrecy level *SECRET*, of categories *EUR* or *US*. *SECRET* is an element of a poset, in fact in this case a total order, which is: *UNCLASSIFIED < CONFIDENTIAL < SECRET < TOPSECRET*. More complex examples are presented in Chin and Older [7], Ch. 13.

The fact that simple labels are sufficient to specify the data flow in networks can be shown in two ways: one is Lemma 2, and the other is the following Property 1.

To simplify the presentation, we consider only composite labels that are couples <*lev, cat*> where *lev* is a secrecy level and *cat* is a set of categories. The reasoning can be generalized to the case of labels with several *lev*s, e.g to express integrity levels also. Composite labels are compared by *tuple comparison* (Sect. 5.2 in Bishop [6]), as follows:

**Definition 6.** Let *lev, lev'* be elements of the same poset and *cat, cat'* be subsets of a set of categories. Then <*lev,cat*> ≤ <*lev',cat'*> iff *lev*≤*lev'* and *cat*⊆*cat'*.

Tuple labels can be transformed into set labels, preserving the partial order:

**Property 1**. Given tuple labels of the form $\lambda$ = <*lev,cat*>, where the values of *lev* are members of the same poset $\Lambda$ and the values of *cat* are subsets of the same set *C*, each $\lambda$ can be translated into a label that is a set of categories, called *Set($\lambda$)*, such that for any two tuple labels $\lambda$ and $\lambda'$, $\lambda \leq \lambda'$ implies *Set($\lambda$)* ⊆ *Set($\lambda'$)*.

*Proof.* Define a set of categories distinct from those in *C*, one for each element of $\Lambda$, and for each *lev*∈$\Lambda$, let *Cat(lev)* be its category. Define: *Set($\lambda$)* = *{Cat(lev'') | lev''≤lev}* ∪ *cat*. To see that $\lambda \leq \lambda'$ implies *Set($\lambda$)*⊆*Set($\lambda'$)*, let $\lambda'$ = <*lev',cat'*>. By Def. 6, $\lambda \leq \lambda'$ implies *lev*≤*lev'*, which implies *{Cat(lev'') | lev''≤ lev}* ⊆ *{Cat(lev'') | lev''≤ lev'}*, also $\lambda \leq \lambda'$ implies *cat*⊆*cat'*.

In the example above, take $\lambda$ =<*SECRET, {EUR, US}*> for which *Set($\lambda$)* = *{UNCLASSIFIED, CONFIDENTIAL, SECRET, EUR, US}*. Note that *UNCLASSIFIED*, etc. in *Set($\lambda$)* are categories for which we have kept the names of the corresponding security levels. Consider a greater tuple label, such as $\lambda'$=<*TOPSECRET, {EUR,US,RUS}*>. Then Set($\lambda'$) = *{UNCLASSIFIED, CONFIDENTIAL, SECRET, TOPSECRET, EUR, US, RUS}*. We have: *Set($\lambda$)*⊆*Set($\lambda'$)* as expected. Hence, tuple labels may be more intuitive and shorter than simple set labels, but they can be translated into equivalent simple set labels, comparable by simple set inclusion. Simple set labels can be more expressive than tuple labels, since they can contain more than one element of a partial order of security levels. To see this, consider a secrecy hierarchy, such as: *UNRESTRICTED < TOPSECRET* and *PUBLIC < CLASSIFIED < TOPSECRET*. With set labels, one can label data in the following manner: *{UNRESTRICTED, PUBLIC}*. This possibility does not exist for couple label as they are normally used, since only one secrecy level, member of a total order, is indicated in them. In [14] we show an example with tuple and set labels more complex than those shown above, expressing both secrecy and integrity levels.

## 7. Coexisting partial orders and intransitive flows

In real networks, many channels exist for different purposes, and if they are all taken to define only one data flow relation, all the entities might collapse into very few equivalence classes, or even a single one, thus making it difficult to identify practically useful secrecy or integrity levels. A method to palliate this problem is well-known in security, it consists in defining different data types, a separate data flow for each type, and *trusted entities* (Bell La Padula [3] Sect. IV) that participate in several flows but are trusted to keep them separate. For example, in banking or in the military there are rules regarding who can tell what to whom. A trusted entity can be understood as a set of distinct parts, with internal rules concerning the data flow among them. This can be considered to establish *intransitive data flows* as in Rushby [15] or, we propose, separate data flows.

In the example of Fig. 1, an additional data flow from *E* to *G* creates the equivalence class *[A,C,E,G]*. This can be avoided if it can be considered that entities *E* and *G* are each split into parts *E', E''* and *G', G''*, that the data flow of Fig. 1 involves *E'* and *G'* and that the new data flow is from *E"* to *G"*, as well that the two data flows concern different data types. For example, in a commercial situation we would have the *order* data flow and the *billing* data flow, each travelling in opposite directions. Companies that receive the orders send out bills, possibly through some of the same intermediary entities, but the billing flow is kept separate from the ordering flow and concerns different parts of the involved entities, or different entities altogether.

Communication among the different parts of a trusted entity must be restricted according to rules that will vary according to the security needs of specific applications. It might be required that the different parts do not communicate at all, or that they can communicate data after transformations that are known in the literature as sanitation, encryption, anonymization, de-identification, purging, etc. These operations result in data type conversions.

## 8. Conclusions

We have shown that some well-known data security concepts are direct application of very basic concepts in the theory or partial orders. These concepts are found in several variations in data security theory. One variation is the *lattice-based data security theory*, by which secure data flows should follow a lattice structure, as determined by the inclusion structure of sets of predefined labels. However the theory presented here is applicable to any network as in Def. 1, and not only to lattice networks, thus generalizing the results of Sandhu [16]. Methods are known to transform arbitrary networks into lattices, but our theory requires no such transformation, and the more constrained lattice concept does not need to be used, see the example of Fig.1.

We have presented the order-theoretical reasons for the relationships existing between the concepts of access control, data flow control and labels, including algorithms to go from one to the others. We have also shown that simple set labels are sufficient to express security data flow control constraints in arbitrary networks. Coexisting partial orders over trusted entities can model coexisting but separate data flows.

A fundamental property in our theory states that data transfer from $x$ to $y$ is allowed iff the label of $x$ is included in the label of $y$. The fact that at any state of an access control or data flow control system, any set of rules can be reduced to simple set inclusion tests, may seem surprising, but is a powerful unifying concept.

While established data security theory is mostly concerned with defining levels of security in networks, by using the concepts presented here it is possible to efficiently determine what are the levels of security implicit in any network. In fact, from the design point of view, Theorems 1 and 2 mean that a security designer could start from any three of the following views (or combinations thereof, if appropriate): desired data flows among entities, desired data flows among groups of entities sharing data, or labels. The other two views follow. Techniques should be developed to translate between labels and established security constructs such as roles, encryption schemes, routing tables, or others as appropriate for the application. These are subjects for future research.

Further discussion on this topic, with other examples, can be found in [14][18].

**Acknowledgment.** The author is indebted to his colleagues Jurek Czyzowicz and Andrzej Pelc as well as to former student Abdelouadoud Stambouli for discussion and research hints.

**Funding.** This research was funded in part by a grant of the Natural Sciences and Engineering Research Council of Canada (NSERC). No conflicts of interest exist for this work.

# References

1. A. V. Aho, M. R. Garey, J. D. Ullman, The transitive reduction of a directed graph. SIAM Journal on Computing, 1(2), 1972, 131-137.
2. J. Bang-Jensen, G.Z.Gutin. *Digraphs – Theory, algorithms and applications.* Springer, 2nd Edition, 2009.
3. D.E. Bell, L.J. La Padula. Secure computer systems: unified exposition and Multics interpretation. MTR-2997, Mitre Corp., Bedford, Mass., 1976.
4. M. Ben-Or. Lower bounds for algebraic computation trees, in: Proc. 15th Annual ACM Symposium on Theory of Computing, 1983, ACM, 80–86.
5. G. Birkhoff. *Lattice Theory,* American Mathematical Society, 1967.
6. M. Bishop. *Computer security, Art and science.* 2nd edition. Addison-Wesley, 2019.
7. S.-K. Chin, S. Older. *Access control, security and trust. A logical approach.* CRC Press, 2011.
8. T. H. Cormen et al. *Introduction to Algorithms*, 4nd ed. MIT Press, 2022.
9. D.E. Denning. A lattice model of secure information flow. Comm. ACM 19(5), 1976, 236-243.
10. R. Fraïssé. *Theory of relations.* North-Holland, 1986.
11. F. Harary, R.Z. Norman, D. Cartwright. *Structural models: an introduction to the theory of directed graphs.* Wiley, 1965.
12. E. Harzheim. *Ordered sets.* Springer, 2005.
13. B.W. Lampson. Protection. Proc. 5th Princeton Conference on Information Sciences and Systems (1971). p. 437.
14. L. Logrippo. Multi-level models for data security in networks and in the Internet of things. Journ. of Inform. Security and Appl. (Elsevier) 58: 102778 (2021).
15. J. Rushby. Noninterference, transitivity and channel-control security policies. Technical Report, SRI International, May 2005.
16. R.S. Sandhu. Lattice-based access control models. IEEE Computer 26(11), 1993, 9–19.
17. R.S. Sandhu. On five definitions of data integrity. In Database Security VII: Status and Prospects, North-Holland, 1994, 257-267.
18. A. Stambouli, L. Logrippo. Data flow analysis from capability lists, with application to RBAC. Information Processing Letters, 141(2019), 30-40.